# A TDM-based Analog Front-End for Ear-EEG Recording with 74.5-GΩ Input Impedance, 384-mV DC Tolerance and 0.27-μV$_{rms}$ Input-Referred Noise

Huiyong Zheng[1,2], Wenning Jiang[2] and Xiao Liu[1,2]
[1]*School of Information Science and Technology, Fudan University*
[2]*State Key Laboratory of Integrated Chips and Systems, Fudan University, Shanghai 201203, China*
Email: xiao@fudan.edu.cn

*Abstract*—This paper presents the design of a time-division multiplexed capacitively-coupled chopper analog front-end (AFE) with a novel impedance boost loop (IBL) and a novel DC servo loop (DSL). The proposed IBL has two impedance booting loops for compensating leakage current due to parasitic capacitance from the ESD pad and external interconnections, and the chopper. By shifting the compensation node from the feedback pathway to the amplifier's inputs, this work realizes a higher-resolution compensation, boosting the input impedance of the AFE to several tens of GΩ. The proposed DSL consists of a coarse DSL driven by DC supply voltages and a fine DSL driven by five phase-interleaving pulse-width modulated waveforms (PI-PWM). Avoiding the usage of ΔΣ CDAC, the energy efficiency is better than conventional DSLs. Designed in a 0.18-μm CMOS process, the AFE consumes 4.5 μA from a 1.2-V supply. The simulated input referred noise is 0.27 μV$_{rms}$ from 0.5 to 100 Hz in the presence of a 384-mV EDO. With a 10-pF parasitic capacitance, the proposed AFE achieves an input impedance of more than 74.5 GΩ at 1 Hz and 6.4 GΩ at 50 Hz. The simulation results have been robust under 100 Monte-Carlo samples.

## I. INTRODUCTION

EEG (Electroencephalogram) has been widely used for monitoring of various neurological disorders such as epilepsy and strokes [1]. However, EEG recording from the scalp takes a long time to set up and the procedure is usually carried out by a trained professional. The recording front-end is typically connected to electrodes via long cables, making it sensitive to motion artifacts. Hence, scalp-EEG recording is not suitable for day-to-day usage outside a laboratory environment [2].

In contrast to scalp-EEG, ear-EEG is recorded in the outer ear or ear canal. Compared to scalp EEG, skins near ear are not covered by hair, offering a simple and comfortable electrode-skin interface. In addition, the ear offers a unique biological structure for hosting a miniaturized recording unit which can be either hidden in the ear canal or hooked onto the ear. Therefore, ear-EEG showcases significant potential for long-term monitoring scenarios. Recently, ear-EEG has found applications in various medical diagnostic fields, including sleep stage classification [3] and hearing threshold test [4]. Additionally, it has been utilized in consumer-oriented brain-computer interfaces for different tasks such as game control, motion imagination [5]

This work was funded by STI 2030 - Major Projects 2022ZD0208900 and NSFC Grant 62150610498.

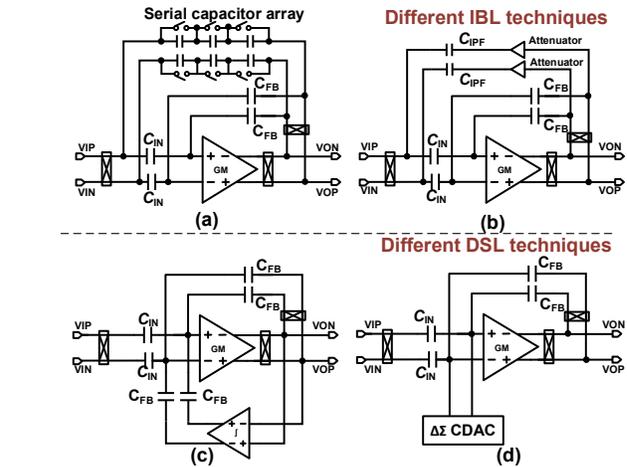

Fig. 1. (a) DSL with an analog-mode integrator. (b) digital DSL with a delta-sigma CDAC. (c) IBL with a serial capacitor array. (d) IBL with an attenuate amplifier.

and self-modulated brain activity classification [6]. However, the ear-EEG signal's amplitude is typically 10-15 dB smaller than that of traditional scalp EEG [1], which increases the requirement of the noise tolerance of the analog front-end (AFE).

For the convenience of daily use, wearable ear-EEG employs dry electrodes [7]. The large electrode impedance of dry electrodes, in the range of a few hundreds of kΩ to a few tens of MΩ [1], leads to the loss of the effective input signal and a decrease in signal to noise ratio (SNR). Conventional recording circuits used positive capacitive feedbacks to boost the input impedance [8]. However, parasitic capacitors associated with the feedback capacitors limit the impedance boost factor (IBF). In order to trim the positive feedback, Ha and his coworkers replaced the fixed feedback capacitor with a serial capacitor array as shown in Fig. 1(a) [9]. However, the improvement to the IBF is still limited by the parasitic capacitance in the switches. In [10], an attenuator amplifier has been employed to reduce the resolution requirement of the capacitor array ($C_{IPF}$) as shown in Fig. 1(b). However, the attenuator amplifier contributes considerable noise to the system.

Time-division multiplexing (TDM) has been widely used in multi-channel recording where area and power are of concern [11]. Multiple channels can share a single AFE with different electrodes. A single amplifier offers an extra benefit of uniform gain, making the gain consistent among individual channels.

Surface electrodes are subject to large electrode offset voltages. In worst case scenarios, the dc-offset voltage of dry electrodes can be up to ±300 mV [12]. To keep the front-end amplifier saturation free, the front-end circuit requires a large dc-suppression capability. Due to the challenge of toggling an integrator with an inherently large time constant across individual channels, analog DC servo loop (DSL) (Fig. 1(c)) and mixed-mode DSL [13, 14] are not suitable for a TDM-based AFE. Although a fully digital DSL may also be slow at the first use, its control word can be stored for future use and recalibration is only necessary when the residual dc voltage exceeds a preset threshold. Shin employed a delta-sigma CDAC to achieve a high-resolution dc suppression [15]. However, the delta-sigma CDAC requires an extra modulator, which increase the power consumption (see Fig. 1(d)).

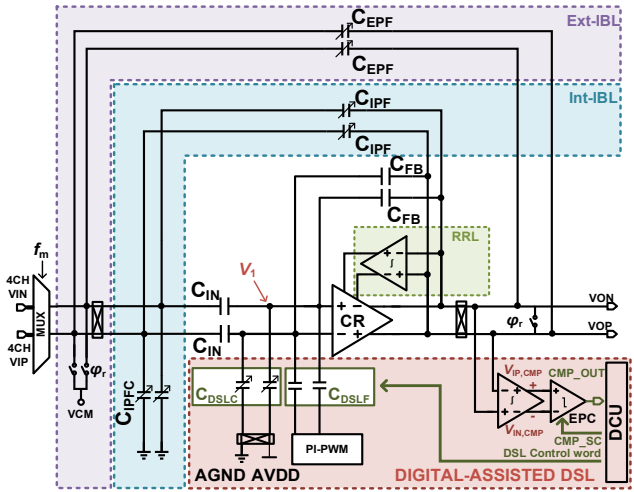

Fig. 2. Architecture of the proposed AFE.

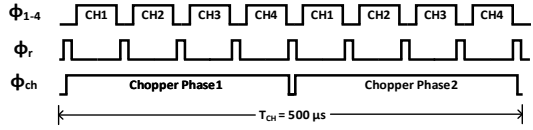

Fig. 3. Timing diagram of the TDM.

This paper proposes a 4-channel TDM-based AFE with high input impedance of up to 74.5 GΩ and a dc tolerance of up to ±384 mV while the input-referred noise is kept below 0.27 µV$_{rms}$. This paper is organized as follows. Section II introduces the system architecture and circuit implementation of the proposed AFE. Section III presents the simulation results. Section IV concludes this work.

## II. CIRCUIT IMPLEMENTATION

### A. System architecture

Fig. 2 shows the proposed ear-EEG AFE which consists of a TDM-based chopper-enabled capacitively-coupled instrumentation amplifier (CCIA); two impedance boost loops (inc. an external IBL, i.e., Ext-IBL, and an internal IBL, i.e., Int-IBL); and a digital-assisted DC servo loop (inc. a coarse DSL and a fine phase-interleaving DSL (PI-DSL)). Both positive and negative inputs of the TDM-CCIA are time-division multiplexed among 4 ear-EEG recording electrode pairs.

The multiplexing frequency, $f_m$, is set to 16 kHz (i.e., 4 kHz per channel which is 2 times higher than the chopper frequency,

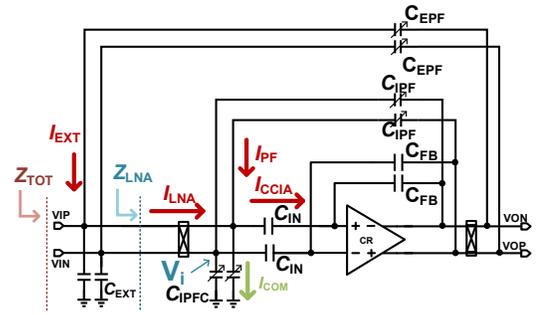

Fig. 4. Proposed IBL.

$f_{ch}$ is 2 kHz). The input capacitors, $C_{IN}$, and feedback capacitors, $C_{FB}$, are set to 40 pF and 0.4 pF respectively, providing a gain of 40 dB. Fig. 3 shows the timing diagram of the TDM control. During φ$_{1-4}$ the AFE selectively amplifies the signals from one of the four channels, while in φ$_r$ the AFE is reset to its optimal common-mode voltages every time when channel switching occurs in order to reduce the crosstalk between successive channels. A current reused transconductance amplifier [13] is employed as the core amplifier to lower the thermal noise.

### B. Two-Stage-Compensation Impedance Boost Loop

Fig. 4 shows the proposed two-stage-compensation IBL. Different from conventional designs, an additional sunk path via $C_{IPFC}$ (see $I_{COM}$ in Fig. 4) has been introduced to provide finer resolution of trimming. The total input impedance of such a circuit looking, $Z_{TOT}$, is given by

$$Z_{TOT} = \frac{1}{sC_{EXT}} \parallel Z_{LNA} \quad (1)$$

where $C_{EXT}$ represents the lumped parasitic capacitance from the ESD pad and external interconnections, and $Z_{LNA}$ is the input impedance of the LNA.

For the sake of analytical simplicity, the contribution of $C_{EXT}$ to the input impedance is initially disregarded. Fig. 4 illustrates all the current paths in the input node. In a convention IBL, the currents into node Vi are composed of $I_{LNA}$, $I_{CCIA}$ and $I_{PF}$. The current flowing into the LNA, $I_{LNA}$, is given by

$$I_{LNA} = I_{CCIA} - I_{PF} = V_I \cdot 2f_{ch} \cdot (C_{IN} - (A_0 - 1) \cdot C_{IPF}) \quad (2)$$

where $I_{CCIA}$ is the current flowing into the CCIA, $I_{PF}$ is the compensation current via $C_{IPF}$, $I_{LNA}$ is the current at the input of the chopper, and $A_0$ is the AFE's closed loop gain. Consequently, an infinite $Z_{LNA}$ can be realized when $I_{PF}$ is equal to $I_{CCIA}$ with $C_{IPF}$ trimmable. The optimal value of the positive feedback capacitor, $C_{IPF,optimal}$, should be equal to $C_{IN}/(A_0-1)$. The improved $Z_{LNA}$ after-trimming can be summarized as

$$Z_{LNA,1} > \frac{1}{2f_{ch} \cdot (A_0 - 1) \cdot \frac{1}{2}C_{IPF,LSB}} \quad (3)$$

where $C_{IPF,LSB}$ is the LSB of the trimming capacitor array. Accordingly, IBF is limited by the minimum cap size of a particular silicon process and the closed-loop gain of the LNA.

Instead of relying on an adjustable $C_{IPF}$ to make $I_{PF} = I_{CCIA}$, the proposed IBL utilizes a fixed $C_{IPF}$ which is intentionally set to slightly larger than the optimal value. A large $C_{IPF}$ causes an over-compensated current flowing into Node $V_i$. But the extra current is sunk by the 6-bit CDAC ($C_{IPFC}$), leading to almost zero $I_{LNA}$. Therefore, the input current of LNA, $I_{LNA}$, is modified as

$$I_{LNA} = I_{CCIA} + I_{COM} - I_{PF} \quad (4)$$

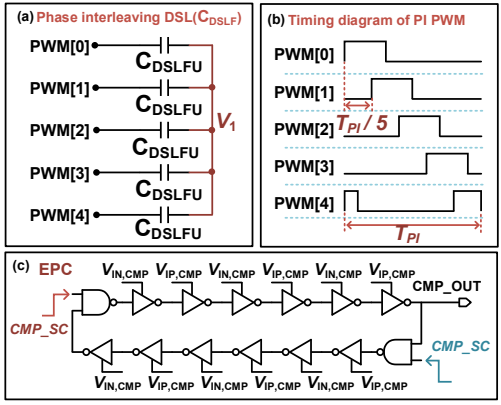

Fig. 5. (a) The CDAC in the fine DSL($C_{DSLF}$). (b) Timing diagram of the PI-PWM. (c) Implementation of the EPC.

Hence $I_{LNA} = V_I \cdot 2f_{ch} \cdot (C_{IN} - (A_0 - 1) \cdot C_{IPF} + C_{IPFC})$ (5)

where $I_{COM}$ is the compensation current via $C_{IPFC}$. The boosted $Z_{LNA}$ is given by,

$$Z_{LNA,2} > \frac{1}{2f_{ch} \cdot \frac{1}{2} C_{IPFC,LSB}} \quad (6)$$

where $C_{IPFC,LSB}$ is the LSB of the trimming capacitor array $C_{IPFC}$. Comparing (3) with (6), $Z_{LNA2}$ can be ($A_0$-1) times larger than $Z_{LNA1}$ when the LSB of $C_{IPFC}$ is equal to that of $C_{IPF}$. In this work, the LSB of $C_{IPFC}$ is designed as 5 fF, which is the minimum MIM capacitor size allowed by the CMOS process.

When $C_{EXT}$ is taken into account, an additional current, $I_{EXT}$, is present. $C_{EXT}$ becomes effectively $C_{EXT}/f_{ch}$ at node Vi (see Fig. 4(a)). Limited by the resolution of $C_{IPFC}$, such a small $C_{EXT}/f_{ch}$ is difficult to compensate by the $C_{IPFC}$ alone [10]. Therefore, we place an external IBL, similar to [10], to form a negative capacitor to compensate for the $C_{EXT}$. The new overall input impedance of the CCIA becomes,

$$Z_{TOT} = \frac{1}{s(A_0 - 1) \cdot \frac{1}{2} C_{EPF,LSB}} \parallel Z_{LNA,2} \quad (7)$$

where $C_{EPF}$ is an adjustable CDAC on the external IBL path, whose LSB ($C_{EPF,LSB}$) is also 5 fF. With $A_0 = 40$ dB and $f_{ch} = 2$ kHz, theoretical calculation reveals that $Z_{TOT}$ at 50 Hz and 1 Hz are larger than 12 GΩ and 100 GΩ, respectively.

### C. DC Servo Loop

The proposed DSL is composed of a coarse DSL and a phase-interleaving DSL (PI-DSL) for fine adjustment (see Fig. 2). The coarse DSL consists of a 7-bit capacitor DAC, $C_{DSLC}$, with an LSB of 50 fF, and it is driven by 1.2-V supply voltage. Although the maximum input-referred EDO that can be suppressed by the coarse DSL is ±384 mV, which is constrained by the size of the LSB capacitor, it may still leave a residual input-referred EDO of up to 3 mV. In this work, the residual input-referred EDO has been further eliminated by the proposed fine PI-DSL.

The fine PI-DSL consists of a uniform-weighted 5-bit CDAC ($C_{DSLF}$) with a 10 fF unit capacitor as shown in Fig.5(a). Unlike

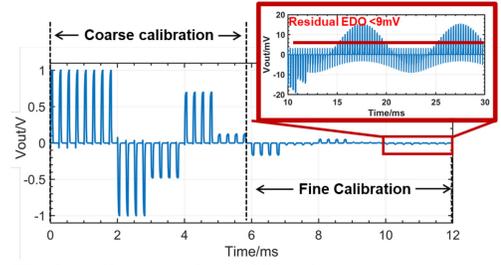

Fig. 6. Timing diagram for calibration

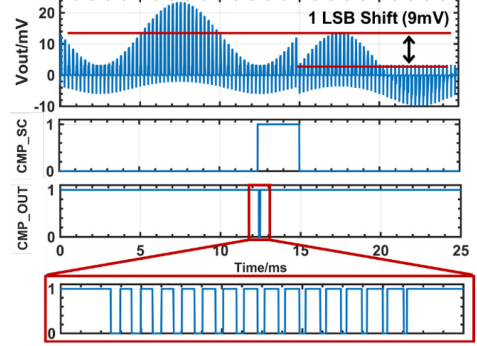

Fig. 7. Timing diagram for monitoring residual EDO.

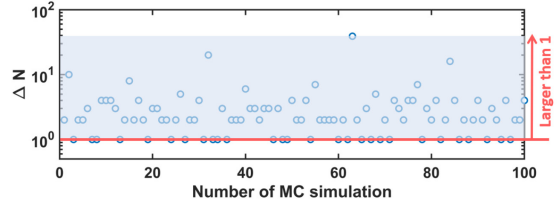

Fig. 8. 100-samples Monte Carlo simulation of ΔN.

$C_{DSLC}$, the $C_{DSLF}$ is driven by five phase-interleaving pulse-width-modulated (PWM) waveforms. All five PWM waveforms exhibit identical pulse width, and each successive PWM waveform holds a phase difference of $2\pi/5$ driving a single bit of $C_{DSLF}$ (see Fig. 5(b)). The period of the PWM, $T_{PI}$, is 1.5 μs and its pulse width is defined by the control words from the digital control unit with a step size of 50 ns. The DC component of the PWM signals suppresses the EDO. Therefore, each LSB of the fine DSL can suppress an EDO of 90 μV. The charges from the five PI-PWM waveforms are summed in $C_{FB}$. The harmonic frequencies of the PI-PWM waveforms are up modulated by a factor of 5, which can easily be removed by the intrinsic low-pass characteristic of the LNA.

Since the EDO changes slowly over time, an integrator with a cut-off frequency of 50 mHz is adequate to monitor the DC component of the LNA's output. An edge-pursuit comparator (EPC) similar to [17] quantizes the residue EDO. If the residual EDO is larger than one LSB of the PI-DSL, one increment or decrement is implemented to the existing code of the PI-DSL according to the EPC's output. Fig. 5(c) shows the implementation of the EPC, which consists of two NAND gates and two 6-stage inverter chains. At the beginning, the START_COMPARE signal (CMP_SC) is low and no power is consumed. Once the CMP_SC goes high, two edges start to progress along the inverter chains. As the progression speed of the two edges are determined by the differential inputs to the

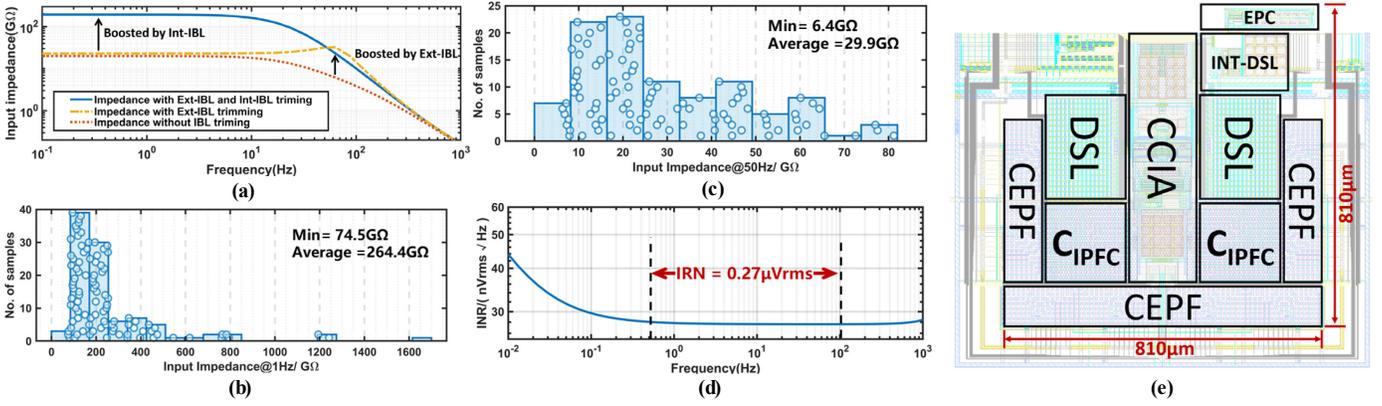

Fig. 9 (a) post-layout simulated input impedance with $C_{EXT}$=10 pF (b-c) 100-samples Monte-Carlo simulation of input impedance with $C_{EXT}$=10 pF (d) Simulated input referred noise. (e) Layout of the proposed AFE.

EPC, i.e., $V_{IP, CMP}$ and $V_{IN, CMP}$ (see Fig. 2), the two edges travel at different speeds until one overtakes the other. The duration required for the EPC to execute a comparison is determined by the amplitude of the EPC's inputs.

Therefore, the EPC inherently possesses a coarse quantization capability, which can be used to quantize and monitor the residual EDO. The relationship between $|V_{IP, CMP} - V_{IN, CMP}|$ and EPC output's (CMP_OUT) is given by (8)

$$N = \frac{I_B}{2g_m V_{I,CMP}} \quad (8)$$

where $N$ is number of oscillation cycles in the CMP_OUT signal during the ON-phase of the CMP_SC, $I_B$ is the bias current of the EPC, $g_m$ is the transconductance of the EPC's input transistor. The code for the PI-DSL changes by one when $N$ is less than a threshold value ($N_{TH}$) which corresponds to the number of oscillation cycles equivalent to an input-referred EDO of 90 μV. As $N_{TH}$ is sensitive to PVT variations, each chip has an identical $N_{TH}$, which was obtained by counting the number of rising edges of the CMP_SC when connecting the input of the AFE to ground and setting the fine-DSL to 1 LSB.

In this work, each channel undergoes a one-time calibration. Subsequently, during the operational mode, the EPC samples the residual DC voltage at the LNA's output every 8 seconds. Only when the measured $N$ is smaller than $N_{TH}$, the fine PI-DSL increases or decreases by one LSB depending on the output of the EPC. Fig. 6 shows the timing diagram of the proposed DSL's calibration when one channel has a large EDO and the inputs of other channels are shorted. A bit shift of the fine-DSL occurs when the residual EDO is larger than the threshold (see Fig. 7).

To evaluate the resolution of the EPC, the number of oscillation cycles of the EPC with different input are simulated. Fig. 8 shows 100-samples Monte Carlo simulation results when the residual voltage at the output of the AFE is equal to 90μV and 135μV. The simulated difference between the number of EPC' oscillations (ΔN) consistently maintain the value of being large than 1, indicating that the resolution of the EPC is higher than 45 μV and the residual DC offset at the AFE's input would be less than 135 μV after the PI-DSL compensation.

### III. SIMULATION RESULT

The proposed AFE has been validated in the Cadence Virtuoso Environment using X-FAB's 0.18-μm CMOS process. The AFE consumes 18-μA total current under a 1.2-V supply. Fig. 9(a) shows the post-layout simulated input impedance of the AFE with $C_{EXT}$ = 10pF, the simulated input impedance is 192 GΩ at 1Hz and 40 GΩ at 50 Hz. Fig. 9(b-c) show the Monte Carlo simulated results (100-samples) with a 10-pF $C_{EXT}$. It's evident that the input impedance is higher than 6.4 GΩ at 50 Hz and 74.5 GΩ at 1 Hz, with an average value of 29.9 GΩ at 50 Hz and 264.4 GΩ at 1 Hz. The slight deviation observed from the theoretical value is due to the mismatch of the CDAC's switch. Fig. 9(d) shows the simulated input-referred noise (IRN) of the AFE. The INR is 0.27 μ$V_{rms}$ from 0.5-100 Hz at the presence of ±384 mV EDO. Fig. 9(e) shows the layout of the AFE. Table I shows the performance summary of the proposed AFE and how it compares to the state-of-the-art. The proposed AFE achieves the highest input impedance and large EDO tolerance while maintaining a competitive input referred noise to the state-of-the-art designs.

Table. I. PERFORMANCE & COMPARISON

|  | JSSC18[13] | JSSC22[10] | JSSC22[15] | **This work[a]** |
|---|---|---|---|---|
| Process | 65nm | 110nm | 65nm | **180nm** |
| Supply(V) | 0.8 | 1 | 1.2 | **1.2** |
| Power/Ch | 4.5μW | 3.8μW | 0.7μW | **5.4μW** |
| INR(μ$V_{rms}$) | 0.38 | 0.36 | 3.2 | **0.27** |
| BW (Hz) | 0.5-100 | 0.5-300 | 1-500 | **0.5-100** |
| NEF | 3.32 | 1.54 | - | **2.2** |
| $Z_{IN}$ (Ω) | - | 15G @10Hz | - | **>74.5G @1Hz** |
|  | 1G @60Hz | 2G @50Hz | 24.5M @100Hz | **>6.4G @50Hz** |
| EDO (mV) | 350 | - | 50 | **384** |

[a] Simulated Result

### IV. CONCLUSION

A TDM-based AFE suitable for ear-EEG recording has been proposed in this paper. Thanks to the novel impedance boost loops and phase-interleaving DC servo loop, the AFE achieves an ultra-high input impedance of 74.5 GΩ and a large EDO tolerance up to ±384 mV.